# Tunable strong plasmon-exciton coupling based on borophene and deep subwavelength perovskite grating

**Xiao-Fei Yan,[1, 2] Qi Lin,[1, 3] Gui-Dong Liu,[1, 3, *] and Ling-Ling Wang [3]**

[1]*School of Physics and Optoelectronics, Xiangtan University, Xiangtan 411105, China*
[2]*Hunan Engineering Laboratory for Microelectronics, Optoelectronics and System on a Chip, Xiangtan University, Xiangtan 411105, China*
[3]*School of Physics and Electronics, Hunan University, Changsha 410082, China*
*gdliu@xtu.edu.cn

**Abstract:** Two-dimensional materials support deeply confined and tunable plasmonic modes, which have great potential for achieving device miniaturization and flexible manipulation. In this paper, we propose a diffraction-unlimited system (period ≈ λ/20) composed of borophene layer and perovskite grating to investigate the strong coupling between the borophene guiding plasmon (BGP) and perovskite exciton (PE) mode. The resonant energy of BGP mode could be electrically tuned to match the energy of PE mode, and a remarkable Rabi splitting is attained under zero-detuning condition. The splitting energy could reach 230 meV due to the strong field enhancement provided by BGP mode. Consequently, an active reflective phase modulation with 1.76π range is achieved by dynamically manipulating the detuning. Furthermore, by increasing the distance between the borophene layer and perovskite grating, a parity-time symmetry breaking could be observed with the vanished energy splitting. Our results deepen the understanding of light-matter interaction at the sub-wavelength scale and provide a guideline for designing active plasmonic devices.

## 1.  Introduction

The interaction between the quantum emitters and trapped optical field in nanostructure has achieved considerable attention and is widely researched in the fundamental physical phenomena and further applications [1, 2]. When the energy exchange rate of the emitters and the optical mode exceeds the respective damping rate, the light-matter interaction would be driven into the strong coupling regime characterized by a remarkable Rabi splitting [3, 4]. And a half-light and half-matter bosonic quasiparticle state is formed, termed exciton-polariton. The mixed state combines the advantage of either the photon or exciton properties, such as robust coherence and low effective mass, which can be investigated in a variety of applications such as Bose-Einstein condensation [5], topological insulator [6] and all-optical switching [7]. Nevertheless, suffering from the small binding energy, the exciton-polariton achieved in conventional semiconductor materials such as cadmium telluride [8] and zinc oxide [9] integrated with the nanostructure system could only be observed at cryogenic temperature. The hybrid inorganic-organic perovskite is used to investigate the strong coupling with the optical modes for its excellent properties, including exciton response at room temperature, tunable broadband band gap, and convenient fabrication [10–14].

A large coupling strength $g$ with low damping rate is the key to accomplishing the strong exciton-polariton coupling and $g \propto \sqrt{N/V}$ in which $N$ and $V$ are exciton number and mode volume, respectively [15, 16]. Though several hybridized systems are designed to reduce the damping rate, such as optical cavity [17] and photonic crystal [18], the mode volume of these systems cannot be suppressed due to the diffraction limit, resulting that the coupling strength is confined. Alternatively, the metal nanostructures support plasmons that confine the optical field at the nanoscale to suppress the mode volume greatly [19, 20]. Nevertheless, the resonant energy of the plasmonic mode can only be passively changed by the geometrical parameter or dielectric environment to cross the resonant energy of the exciton mode in the plasmon-exciton coupling process [21, 22]. The highly confined plasmons in 2D materials, such as graphene or black phosphorus, could be electrically tuned the resonant wavelength [23, 24]. However, the wavelengths of plasmons in these 2D materials (mid-infrared and far-infrared region) are mismatched with the wavelength of PE mode in the visible region. In other words, the strong coupling between these plasmonic modes with PE mode is difficult. Fortunately, borophene, as an emergent 2D material, supports the plasmonic mode in the visible region owing to its high electron density (~$10^{19}$ m$^{-2}$) [25]. Moreover, the borophene plasmon with extrasmall mode volume could be flexibly controlled to manipulate the plasmon-exciton hybridized process [26, 27]. Combined with the borophene and perovskite, the tunable strong plasmon-exciton coupling will be a desired platform to investigate the light-matter interaction at room temperature.

In this work, a hybrid system supported in borophene-assisted deep subwavelength perovskite grating (period ≈ $\lambda$/20) is proposed to investigate the strong plasmon-exciton coupling between BGP and PE mode. The resonant energy of the BGP mode is actively tuned to match that of PE mode, and a Rabi splitting can be observed under the zero-detuning condition. The Rabi splitting energy is up to 230 meV due to the strong field enhancement offered by the BGP mode. Furthermore, by electrically manipulating the electron density of borophene, the system shows a 1.76π modulation range, which could be attributed to detuning between BGP and PE mode altered from 0 (coupled) to 186 meV (decoupled). Moreover, a parity-time symmetry breaking could be observed when the distance between borophene layer and perovskite grating increases. As a result, the energy splitting in the absorption spectra would vanish, and the linewidth of the hybridized mode undergoes a conversion from non-degeneracy to degeneracy.

## 2. Design and simulations

The proposed nanostructure consists of a continuous borophene layer and an etched perovskite grating, as depicted in Fig. 1(a). The optimized geometrical parameters $P$ (period), $w$ (width), and $h$ (height) are set as 20, 10, and 20 nm, respectively. The permittivity of the borophene is calculated via [26],

$$\varepsilon_{r,jj} = \varepsilon_d - \frac{e^2 n}{m_j \varepsilon_0 t (E^2 + \frac{1}{\tau^2})}, \varepsilon_{i,jj} = \frac{e^2 n/\tau}{m_j \varepsilon_0 t (E^2 + \frac{1}{\tau^2})}, \quad (1)$$

in which $\varepsilon_d$ = 11 and $t$ = 0.3 nm are the direct current permittivity of boron and thickness of the borophene layer, respectively. $m_j$ ($j = x$ or $y$) is effective electron mass of optical axes in which $m_x$ = 1.4 $m_0$ and $m_y$ = 3.4 $m_0$. $\tau$ = 65 fs is electron relaxation time. $e$ is electron charge, and $n$ is electron density. The organic-inorganic 2D layered perovskite (C$_6$H$_5$C$_2$H$_4$NH$_3$)$_2$PbI$_4$ is used to investigate the plasmon-exciton coupling in this work. The permittivity of the perovskite is given as follows [11],

$$\varepsilon(E) = n_p^2 + \frac{A}{E_{exciton} - E^2 - i\gamma E}, \qquad (2)$$

where $n_p$ = 2.15 is the background index, $E_{exciton}$= 2.394 eV is the exciton energy, $A$ =0.48 eV$^2$ is constant related with the oscillator strength and $\gamma$ = 0.03 eV is the homogeneous linewidth. The numerical results are obtained from the finite-difference time-domain (FDTD) method. The periodic boundary condition is set in $x$-direction and perfectly matched layers (PMLs) boundary in $z$-direction. The $x$-direction polarized plane waves are incident in $z$-direction.

## 3. Results and discussion

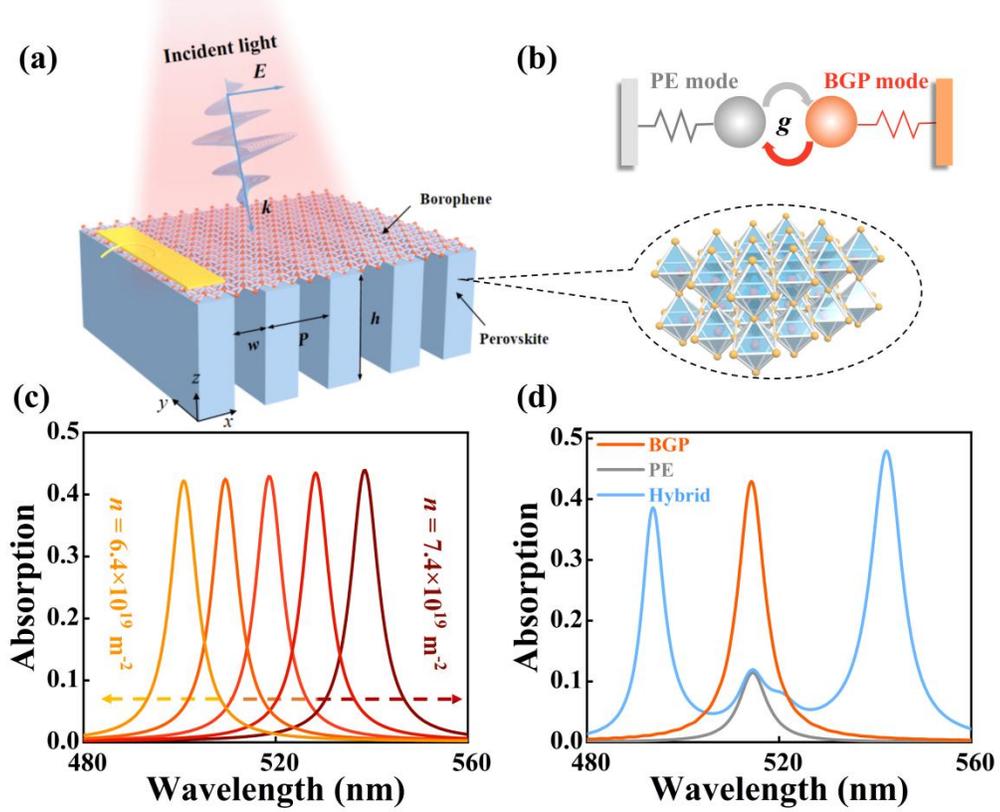

Fig. 1. (a) Schematic of the proposed plasmon-exciton hybridized system. (b) Theoretical coupled oscillator model. (c) The absorption spectra as a function of varied electron density $n$. (d) The absorption spectra of the borophene guiding plasmon, perovskite exciton, and hybridized modes.

To illustrate the physical mechanism of the BGP mode, the imaginary part of the $\varepsilon_{perovskite}$ is set as zero without regard to the exciton mode. In this case, the permittivity of the perovskite grating is $\varepsilon = n_p^2$, and the grating can compensate the wave-vector for exciting the BGP mode in the continuous borophene layer. As demonstrated in Fig.1(c), the resonance wavelength of the BGP mode is actively tuned by altering the electron density $n$. The resonant energy of the BGP mode (orange line) could be equal to that of the PE mode (grey line) by tuning the electron density of borophene layer. An obvious Rabi splitting with two energy levels (2.52 eV and 2.29 eV) occurs when the energy exchange rate exceeds the damping rate in the zero-detuning condition. The Rabi splitting energy reaches 230 meV, benefiting from the enhanced coupling strength by introducing the highly confined plasmon mode in the coupling process. The dispersion curves of the hybrid modes are illustrated in Fig. 2(a). The PE mode remains unchanged (grey line) while the BGP mode is blue-shifted (blue line) with the increase of the electron density. Due to the strong coupling between the PE and BGP mode, a markedly anti-crossing behavior could be observed with the upper polariton (UP) band and lower polariton (LP) band, respectively. The process could be theoretically analyzed via the coupled oscillator model [12, 28]

$$\begin{bmatrix} E_B + i\gamma_B & g \\ g & E_P + i\gamma_P \end{bmatrix} \begin{pmatrix} \alpha_B \\ \alpha_P \end{pmatrix} = E_{UP,LP} \begin{pmatrix} \alpha_B \\ \alpha_P \end{pmatrix}, \quad (3)$$

where $E_B$ ($E_P$) and $\gamma_B$ ($\gamma_P$) denote the resonant energy and damping rate of the original BGP (PE) mode, and $g$ is the coupling strength. $\alpha_B$ and $\alpha_P$ is the mixing fraction representing the weighting of the BGP and PE mode in the UP (LP) band and satisfy the condition of $|\alpha_P|^2 + |\alpha_B|^2 = 1$. By solving Eq. (1), the complex eigenvalues $E_{UP}$ and $E_{LP}$ of the hybridized system are given as follows

$$E_{UP,LP} = \frac{E_B + E_P}{2} \pm \frac{1}{2}\sqrt{4g^2 + [\delta - i(\gamma_B - \gamma_P)]^2} + i\frac{\gamma_B + \gamma_P}{2}. \quad (4)$$

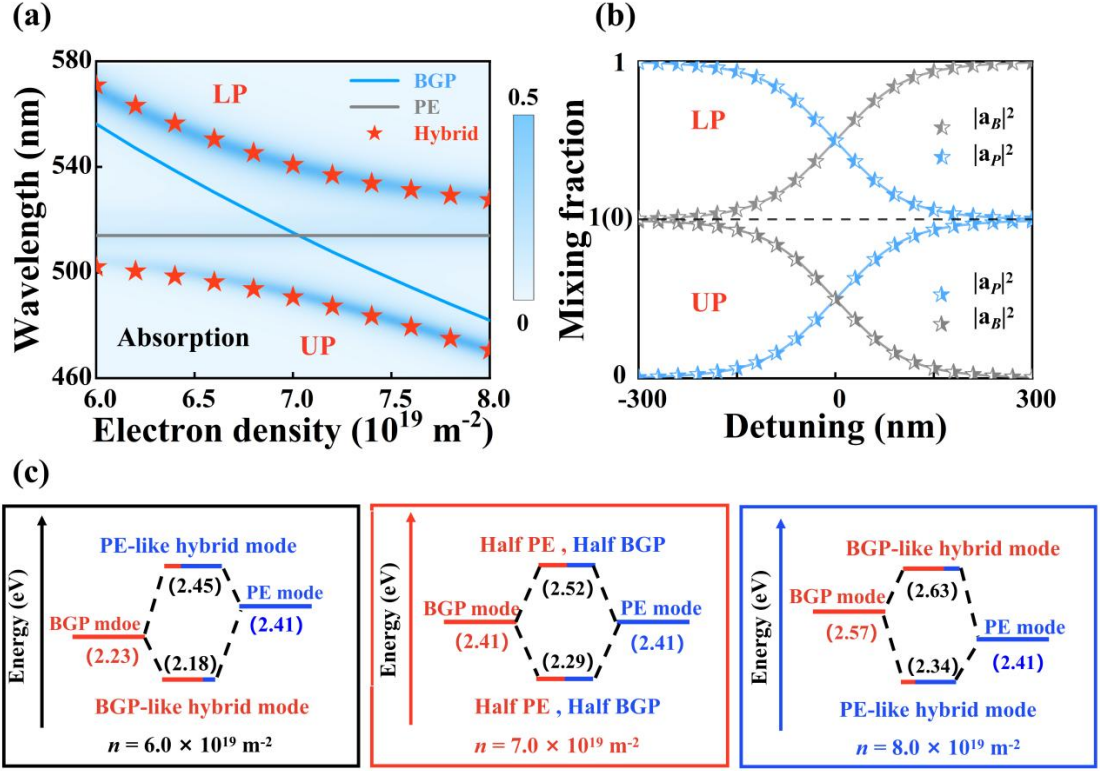

Fig. 2. (a) The hybridized modes dispersion curves with a varied electron density $n$ and theoretically analyzed hybrid modes (stars), original BGP mode (blue line), and PE mode (grey line) mode. (b) The mixing fraction of the upper and lower dispersion curves as a function of detuning. (c) The energy schematic cases in Fig. 2(a).

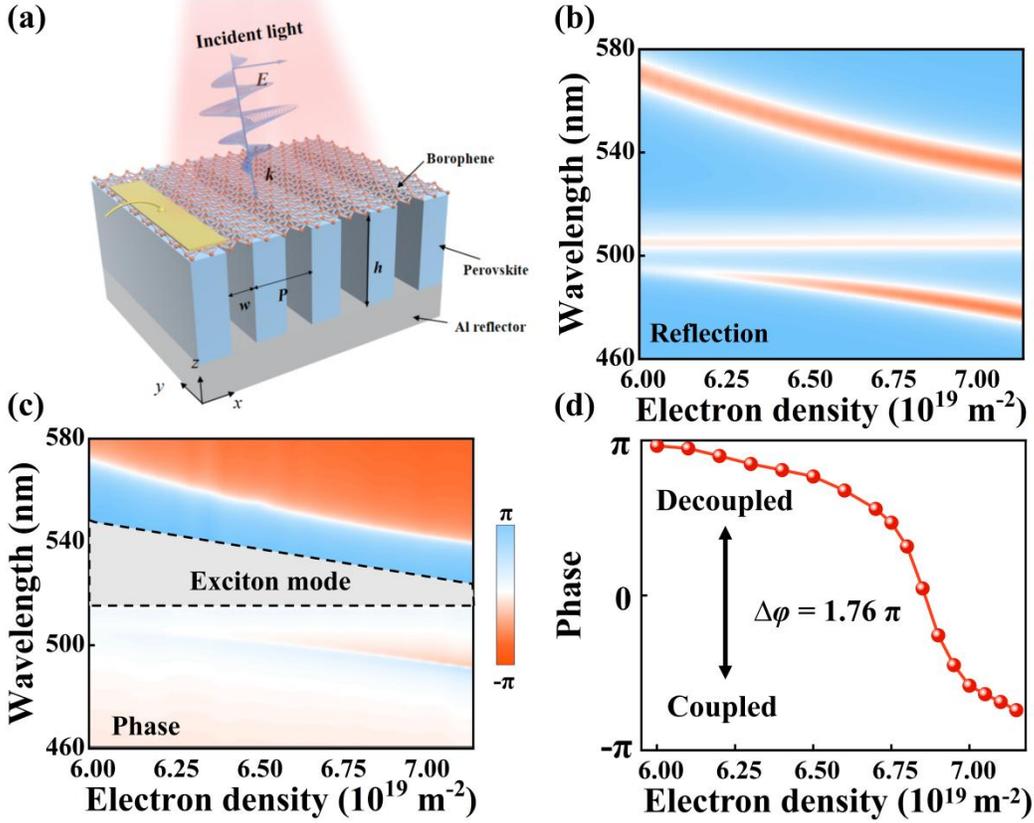

Fig. 3. (a) The schematic of the phase modulator. The reflection (b) and phase (c) dispersion curves as a function of the electron density $n$. (d) Calculated phase at 545 nm with a varied electron density $n$.

In the zero-detuning condition ($\delta = E_B - E_P = 0$), Eq. (2) can be deduced as

$$E_{UP,LP} = \frac{E_B + E_P}{2} \pm \frac{1}{2}\sqrt{4g^2 - (\gamma_B - \gamma_P)^2} + i\frac{\gamma_B + \gamma_P}{2}. \quad (5)$$

The coupling strength can be obtained by checking the Rabi splitting energy and the damping rates of the original modes as

$$g = \frac{1}{2}\sqrt{\Omega^2 + (\gamma_B - \gamma_P)^2}. \quad (6)$$

By calculating the full width at half maximum (FWHM) of the absorption spectra in Fig. 1(d), the damping rates of the original BGP and PE mode are 15.7 and 16.3 meV, respectively. According to Eq. (6), the coupling strength is 115 meV and satisfies the following conditions,

$$g > \frac{|\gamma_B - \gamma_P|}{2}, \quad \text{and} \quad g > \frac{\sqrt{(\gamma_B)^2 + (\gamma_P)^2}}{4}, \quad (7)$$

which indicates that the hybridized system is in the strong coupling regime. According to Eq. (5), the resonant energy of the hybridized modes could be theoretically predicted (red stars) shown in Fig. 2(a) and are accord with the simulation results. The mixing fraction of the UP and LP band could be deduced from Eq. (3),

$$|\alpha_B|^2 = \frac{1}{2}\left(1 \pm \frac{\delta}{\sqrt{\delta^2 + 4g^2}}\right), |\alpha_P|^2 = \frac{1}{2}\left(1 \mp \frac{\delta}{\sqrt{\delta^2 + 4g^2}}\right), \quad (8)$$

and depicted in Fig. 2(b). The weighting in the hybridized process varies with the detuning. The energy schematic cases in Fig. 2(c) are given for a better comprehension of the coupling. By comparing the weighting of the original modes calculated via Eq. (8), if the $|\alpha_P|^2$ is larger than the $|\alpha_B|^2$, the hybrid mode can be considered the PE-like mode. Conversely, it is BGP-like mode. It should be noted that, in the zero-detuning condition, the weighting $|\alpha_P|^2$ is equal to the $|\alpha_B|^2$, and the hybridized modes are considered as a half-PE and half-BGP mode

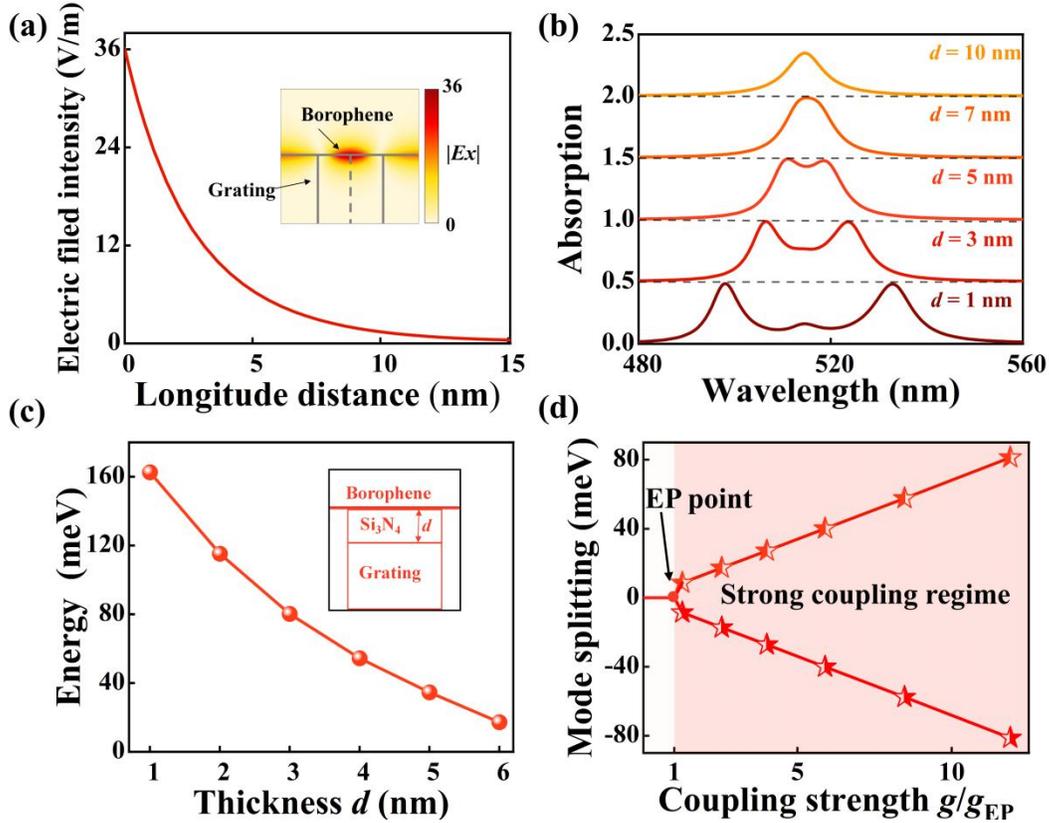

Fig. 4. (a) The normalized electric field intensity fetched from the grey dotted line in the insert as a function of longitude distance. (b) The absorption spectra evolutionary process with the increase of dielectric layer thickness $d$. (c) The splitting energy as a function of dielectric layer thickness $d$. The insert is the side view of the structure. (d) Calculated mode splitting with varied coupling strength.

Based on the electrically tunable plasmon-exciton hybridized system, active reflective phase modulation is accomplished by altering the electron density $n$. As shown in Fig. 4(a), the structure parameters are set the same as that in Fig. 1(a), except that the height $h$ of the perovskite grating is 80 nm. An aluminum layer is used to block the transmission light and enlarge the phase range from 0-π to 0-2π. The optical constant of aluminum is from Palik's book, and the thickness of aluminum is 100 nm [29]. The reflection and phase dispersion curves with the varied electron density $n$ from $6.00 \times 10^{19}$ m$^{-2}$ to $7.15 \times 10^{19}$ m$^{-2}$ are shown in Figs. 3(b) and 3(c), corresponding to the tuning of BGP and PE mode turned from 186 to 0 meV. The reflective phases at 545 nm extracted from Fig. 3(c) are depicted in Fig. 3(d). With the two resonant energy gradual crossing each other, the reflective phase $\varphi$ could be modulated, ranging from 0 to 1.76π.

A parity-time symmetry breaking could be observed in this hybridized system by tuning the coupling strength. The separation distance between the borophene layer and perovskite grating is the key to adjusting the coupling strength. The absolute value of the $x$ component of the electric field in BGP mode is shown in Fig. 4(a) and it exponentially decays with the increased lengthways distance. By introducing silicon nitride dielectric grating with varied thickness between the continuous borophene layer and perovskite grating, the degree of electric field overlap could be controlled and the coupling strength. The index, period and duty cycle of silicon nitride dielectric grating are set as 2.03, 20 nm, and 0.5. The absorption spectra with varied thickness of dielectric grating are shown in Fig. 4(b), and the splitting energy which is confirmed by checking the resonant energy of two peaks are shown in Fig. 4(c). The coupling strength calculated by Eq. (6) decrease from 81 to 8.6 meV with thickness $d$ increased from 1 to 6 nm. When thickness $d$ increases further, the mode splitting vanishes, and the coupling falls into the weak coupling regime. The strong and weak coupling regimes are divided by the exceptional point corresponding to the coupling strength $g_{EP} = |\gamma_B - \gamma_P|/2$. As demonstrated in Fig. 4(d), when coupling strength $g$ is over $g_{EP}$, the system falls into the strong coupling regime characterized by the energy splitting of the hybridized mode. On the contrary, the system falls into the weak coupling regime with the non-degenerate linewidth.

## 4. Conclusion

In conclusion, a tunable strong plasmon-exciton coupling system based on borophene and deep subwavelength perovskite grating is proposed. By actively altering the resonant energy of the BGP mode across the PE mode, an anti-crossing behavior occurs, and the Rabi splitting energy reaches 230 meV due to enhanced electric confinement offered by BGP mode. Based on the dynamical coupling process, a dynamic reflective phase modulation method is proposed and the modulation range is from 0 to 1.76π. Moreover, by increasing the separation distance between the borophene and perovskite grating, the energy splitting would vanish indicating that the system undergoes a parity-time symmetry breaking. The proposed system based on perovskite and borophene materials provides a guideline for designing active plasmonic devices.

**Funding.** National Natural Science Foundation of China (11947062) and Hunan Provincial Natural Science Foundation of China (2020JJ5551, 2021JJ40523).

**Disclosures.** The authors declare no conflicts of interest.

**Data availability.** Data underlying the results presented in this paper are not publicly available at this time but maybe obtained from the authors upon reasonable request.